\documentclass[12pt]{amsart}
\usepackage{a4wide,color}
\usepackage{amsmath,amssymb}
\usepackage[active]{srcltx}

\allowdisplaybreaks

\let\pa\partial

\let\eps\varepsilon

\newcommand{\R}{{\mathbb R}}

\newtheorem{theorem}{Theorem}

\newtheorem{proposition}[theorem]{Proposition}

\newtheorem{corollary}[theorem]{Corollary}

\let\ga=\gamma
\let\de=\delta
\let\eps=\varepsilon

\let\pa=\partial

\begin{document}

\title[Landau equation]{Global in Time Estimates for the Spatially Homogeneous Landau Equation with Soft Potentials}

\author[K.-C. Wu]{Kung-Chien Wu}
\address{Kung-Chien Wu, Department of Pure Mathematics and Mathematical Statistics, University of Cambridge, Wilberforce Road, Cambridge, CB3, 0WB, UK}
\email{kcw28@dpmms.cam.ac.uk; kungchienwu@gmail.com}

\date{\today}

\thanks{It is a pleasure to thank Cl\'{e}ment Mouhot for stimulating discussion concerning this
paper. This work is supported by the
Tsz-Tza Foundation in Academia Sinica (Taiwan) and ERC grant MATKIT. Part of this work was written during the stay
at Institute of Mathematics, Academia Sinica and Department of Mathematics, Stanford University; the author thanks Tai-Ping Liu for his kind hospitality.}

\begin{abstract}
This paper deals with some global in time a priori estimates of the spatially homogeneous Landau equation for soft potentials $\ga\in[-2,0)$. For the first result, we obtain the estimate of weak solutions in $L^{\alpha}_{t}L_{v}^{3-\eps}$ for $\alpha=\frac{2(3-\eps)}{3(2-\eps)}$ and $0<\eps<1$, which is an improvement over estimates by Fournier-Guerin \cite{[Fournier]}. Foe the second result, we have the estimate of weak solutions in $L_{t}^{\infty}L^{p}_{v}$, $p>1$, which extends part of results by Fournier-Guerin \cite{[Fournier]} and Alexandre-Liao-Lin \cite{[Alex]}. As an application, we deduce some global well-posedness results for $\ga\in [-2,0)$. Our estimates include the critical case $\ga=-2$, which is the key point in this paper.
\end{abstract}



\maketitle

\section{Introduction}\label{introduction}
\subsection{The Landau equations}
We consider the spatially homogeneous Landau equation in dimension three for soft potentials.
This equation of kinetic physics, also called Fokker-Planck-Landau equation, has been derived
from the Boltzmann equation when the grazing collisions prevail in the gas. It describes the evolution of the
density function $f_{t}(v)$ of particles having the velocity $v\in\R^{3}$ at time $t\geq 0$:
\begin{align}\label{landau}
\frac{\pa f_{t}(v)}{\pa t}=Q(f_{t},f_{t})(v)\,,
\end{align}
with collision operator
$$
Q(f_{t},f_{t})(v )=\nabla_{v }\cdot\Big\{ \int a(v -v _{*})
\big[f_{t}(v_{*})\nabla f_{t}(v)-f_{t}(v)\nabla_{*} f_{t}(v_{*})\big]dv _{*}\Big\}\,,
$$
where $a(v)$ is a symmetric non-negative matrix, depending on a parameter $\ga\in[-2,0)$,
$$a(v )=|v |^{\ga+2}\mathbf{P}(v )\,,$$
and $\mathbf{P}(v )$ is the $3$ by $3$ matrix
$$
\mathbf{P}(v )=I_{3}-\frac{v \otimes v }{|v |^{2}}\,.
$$

This leads to the usual classification in terms of hard potentials $\ga>0$, Maxwellian molecules $\ga=0$, soft potentials $\ga\in[-2,0)$, very soft potentials $\ga\in(-3,-2)$ and Coulomb potential $\ga=-3$. Just as for the Boltzmann equation, little is known for soft potentials, i.e. $\ga<0$, and even less for very
soft potentials, i.e. $\ga<-2$. In particular, $\ga=-3$ corresponds to the important
Coulombic interaction in plasma physics. Unfortunately, it is also the
most difficult case to study. However, the Landau equation can be derived from the Boltzmann
equation with $\ga\in (-3,-1)$. Note the fact that the more $\ga$ is negative, the more the Landau equation is physically interesting. In this paper, we focus on soft potentials $\ga\in[-2,0)$.

For a given non-negative initial data $f_{{\rm in}}(v)$, we shall use the notations
\begin{align*}
m(f_{{\rm in}})= \int f_{{\rm in}}(v)dv\,,\quad e(f_{{\rm in}})=\frac{1}{2} \int f_{{\rm in}}(v)|v|^{2}dv\,,\quad
H(f_{{\rm in}})= \int f_{{\rm in}}(v)\log f_{{\rm in}}(v)dv\,,
\end{align*}
for the initial mass, energy and entropy. It is classical that if $f_{{\rm in}}(v)\geq 0$ and $m(f_{{\rm in}}), e(f_{{\rm in}}), H(f_{{\rm in}})$ are finite, then $f_{{\rm in}}(v)$ belongs to
\begin{align*}
L\log L(\R^{3})=\Big\{ f(v)\in L^{1}(\R^{3}): \int |f(v)||\log( |f(v)|)|dv<\infty  \Big\}\,.
\end{align*}
The solution of the Landau equation (\ref{landau}) satisfies, at least formally, the conservation of mass, momentum and kinetic energy, that is, for any $t\geq 0$,
\begin{align}\label{conserve}
 \int  f_{t}(v)\varphi(v)dv= \int  f_{{\rm in}}(v)\varphi(v)dv\,,\quad \hbox{for}\quad \varphi(v)=1,v,|v|^{2}\,.
\end{align}
Especially, we define
\begin{align*}
m(f_{t})= \int f_{t}(v)dv\,,\quad e(f_{t})=\frac{1}{2} \int f_{t}(v)|v|^{2}dv\,,\quad
H(f_{t})= \int f_{t}(v)\log f_{t}(v)dv\,.
\end{align*}
Another fundamental a priori estimate is the entropy estimate, that is, the solution satisfies, at least formally, for any $t>0$,
\begin{align}\label{entropy}
\frac{d}{dt}H(f_{t})+\mathcal{D}(f_{t})=0\,,
\end{align}
where
\begin{align}\label{diffusion}
\mathcal{D}(f_{t})=2 \int  \int  dvdv_{*}a(v-v_{*})(\nabla-\nabla_{*})\sqrt{f_{t}(v)f_{t}(v_{*})}(\nabla-\nabla_{*})\sqrt{f_{t}(v)f_{t}(v_{*})}\,.
\end{align}
Note that $a(v-v_{*})$ is a non-negative matrix, this implies $\mathcal{D}(f_{t})\geq 0$, and hence the entropy is decreasing
\begin{align*}
 \int f_{t}(v)\log f_{t}(v)dv\leq  \int f_{{\rm in}}(v)\log f_{{\rm in}}(v)dv\,.
\end{align*}
Foe further use, we define the moment of order $s$, for $s\geq 0$,
$$
\|f_{t}\|_{L^{1}_{s}(\R^{3})}= \int |f_{t}(v)|\big<v\big>^{s} dv\equiv M_{s}(f_{t})\,.
$$
where $\big<v\big>\equiv (1+|v|^{2})^{1/2}$.
And we set
$$
b_{i}(z)=\pa_{j}a_{ij}(z), \quad c(z)=\pa_{ij}a_{ij}(z)\,,
$$
$$
\overline{a}^{f_{t}}_{ij}=a_{ij}*f_{t}\,, \quad \overline{b}^{f_{t}}_{i}=b_{i}*f_{t}\,, \quad \overline{c}^{f_{t}}=c*f_{t}\,.
$$
If $\ga>-3$, we have
$$
a_{ij}(z)=\mathbf{P}_{ij}(z)|z|^{\ga+2},\quad b_{i}(z)=-2|z|^{\ga+2}\frac{z_{i}}{|z|^{2}}, \quad c(z)=-2(\ga+3)|z|^{\ga}\,.
$$
Finally, let us introduce the weak formulation of (\ref{landau}): for any test function $\varphi:\R^{3}\rightarrow \R$,
\begin{align*}
 \int \varphi(v)f_{T}(v)dv= \int \varphi(v)f_{{\rm in}}(v)dv+\int_{0}^{T}dt \int  \int f_{t}(v)f_{t}(v_{*})L\varphi(v,v_{*})dvdv_{*}\,,
\end{align*}
where the operator $L$ is defined by
\begin{align*}
L\varphi(v,v_{*})=\frac{1}{2}\sum_{i,j=1}^{3}a_{ij}(v-v_{*})\pa_{ij}^{2}\varphi(v)+\sum_{i=1}^{3}b_{i}(v-v_{*})\pa_{i}\varphi(v)\,.
\end{align*}

\subsection{Review of previous works}

The theory of the spatially homogeneous Landau equation for hard potentials was studied in great details by
Desvillettes-Villani \cite{[5], [51]}, while the particular case of Maxwellian molecules $\ga=0$ can be found
in Villani \cite{[Villani1]}.

However, to our knowledge, the results concerning about soft potentials may not rich. By using a probabilistic approach,
Guerin \cite{[Guerin]} studied the existence of a measure solution for $ \ga\in(-1,0)$. Still by probabilistic
approach, Fournier-Guerin \cite{[Fournier]} studied the uniqueness of such weak solutions
for soft potentials, moreover, they have global existence result for $\ga\in (-2,0)$ and local existence result for $\ga\in (-3,-2]$. For the Coulomb potential case $\ga=-3$, Arse\'nev-Peskov \cite{[Arsenev]} studied the local existence of weak solutions and Fournier \cite{[Fournier1]} considered the local well-posedness result for such solutions. For the $L^{2}$ a priori estimate, Alexandre-Liao-Lin \cite{[Alex]} constructed the global estimates for $\ga\in(-3,0)$, but their result must be in weighted space and needs smallness assumption of the initial condition for $\ga\in [-3,-2]$.

All these results listed above give a priori estimates of solutions in some $L^{p}$ spaces, globally if $\ga\in (-2,0)$, either locally or globally but in weighted space and need smallness assumption of the initial condition if $\ga\in [-3,-2]$.

In this paper, we construct some global in time a priori estimates for the spatially homogeneous Landau equations, it includes the critical case $\ga=-2$. Moreover, our results do not need in weighted space and smallness assumption.

For simplicity of notations, hereafter, we abbreviate ``{ $\leq C$} " to ``{ $ \lesssim$ }", where $C$ is a positive constant depending only on fixed number.

\subsection{Main result I}
We establish $L^{\alpha}([0,T], L^{3-\eps})$ a priori estimate of the spatially homogeneous Landau equation (\ref{landau}) for $\ga\in [-2,0)$, here $\alpha>1$ and $0<\eps<1$, which is an improvement over estimates by Fournier-Guerin \cite{[Fournier]} in proposition 10 [N. Fournier; H. Guerin, Well-posedness of the spatially homogeneous Landau equation for soft potentials. J. Funct. Anal. 25(2009), no. 8, 2542--2560]. In Fournier-Guerin's work, they obtained global (in time) a priori estimate of weak solutions in $L^{1}_{t}L_{v}^{3-\eps}$ for $\ga\in(-2,0)$. However, our result has the following improvements:
\\
(i) We have better time integrability $L^{\alpha}_{t}$ $(\alpha>1)$.
\\
(ii) Our estimate includes the critical case $\ga=-2$.

\begin{theorem}\label{theorem3}
For any $\ga\in [-2,0)$, $0<\eps<1$ and $s>0$. Let us consider a weak solution $f_{t}(v)$ of \noindent{\rm(\ref{landau})} with initial data $f_{{\rm in}}(v)\in L^{1}_{q}(\R^{3})\cap L\log L(\R^{3})$ for some $q=\frac{-3(\ga-s)(2-\eps)}{\eps}>0$. Then, at least formally, $f_{t}(v)\in L^{\alpha}([0,T],L^{3-\eps})$ with $\alpha=\frac{2(3-\eps)}{3(2-\eps)}$. More precisely,
\begin{align*}
\int_{0}^{T}\|f_{t}\|^{\alpha}_{L^{3-\eps}}dt\lesssim C(T)\,,
\end{align*}
where
\begin{equation}\label{cons}
C(T)\lesssim\left\{\begin{array}{l}
\displaystyle (1+T)^{1+\frac{2\eps(3+\eps)}{3(2-\eps)(2+\ga)}}\,,\quad \ga\in(-2,0)\,,
\\ \\
\displaystyle \exp\big\{ C T^{z}  \big\}\,,\quad z=\frac{(3-\eps)[ 3(2+s)(2-\eps)-2\eps ]}{3(1-\eps)}\,,\quad \ga=-2\,,
\end{array}
\right.\end{equation}
for some constant $C>0$.
\end{theorem}
The proof of this theorem is based on the entropy production estimate, it was firstly introduced
by Desvillettes-Villani \cite{[5]} for hard potentials $\ga>0$, and later Fournier-Guerin \cite{[Fournier]} applied to  $L^{1}([0,T], L^{3-\eps})$ estimate for soft potentials $\ga\in (-2,0)$. In this paper, We can control $\|f_{t}\|_{L^{3-\eps}}$ by using the Hardy-Littlewood-Sobolev inequality for $\ga\in (-2,0)$. However, $\ga=-2$ is the critical case, we apply uniform bound of entropy to get a parameter of freedom, this can help us overcome this difficulty.

\subsection{Main result II}
The second result of this paper is devoted to $L^{\infty}([0,T], L^{p})$ a priori estimate of the spatially homogeneous Landau equation (\ref{landau}) for $\ga\in[-2,0)$, $1<p<\infty$, which extends part of results by Fournier-Guerin \cite{[Fournier]} in proposition 11 [N. Fournier; H. Guerin, Well-posedness of the spatially homogeneous Landau equation for soft potentials. J. Funct. Anal. 25(2009), no. 8, 2542--2560] and Alexandre-Liao-Lin \cite{[Alex]} [R. Alexandre, J. Liao, and C. Lin, Some a priori estimates for the homogeneous Landau equation with soft potentials, arXiv:1302.1814].

In Alexandre-Liao-Lin's work \cite{[Alex]}, they constructed a global (in time) $L^{\infty}([0,T], L^{2})$ a priori estimate for $\ga\in(-3,0)$, but when $\ga\in(-3,-2]$, their estimate needs in weighted space and extra smallness assumption of the initial condition. On the other hand, in Fournier-Guerin's work \cite{[Fournier]}, they have local (in time) $L^{\infty}([0,T], L^{p})$ solutions for $\ga\in(-3, -2]$. However, in this paper, we have global (in time) $L^{\infty}([0,T], L^{p})$ a priori estimate for all $1<p<\infty$, the weighted space and smallness assumption of the initial condition are not necessary in my result.
\begin{theorem}\label{theorem1}
For any $\ga\in[-2,0)$, $1<p<\infty$. Let us consider a weak solution $f_{t}(v)$ of \noindent{\rm(\ref{landau})} with initial data $f_{{\rm in}}(v)\in L^{1}_{2}(\R^{3})\cap L^{p}(\R^{3})$ for $\ga\in (-2,0)$ and $f_{{\rm in}}(v)\in L^{1}_{q}(\R^{3})\cap L^{p}(\R^{3})$ for $\ga=-2$, where $q$ was defined in theorem \ref{theorem3}. Then, at least formally, $f_{t}(v)\in L^{\infty}([0,T],L^{p})$. More precisely,
\begin{align*}
\|f_{t}\|^{p}_{L^{p}}\lesssim C(t)\,,
\end{align*}
where $C(t)\lesssim e^{C t}$ for $\ga\in(-2,0)$ and $C(t)\lesssim \exp\{C t^{z}\}$ for $\ga=-2$, $C>0$ is a positive constant and the polynomial power $z$ was defined in \rm{(\ref{cons})}.
\end{theorem}
The proof of this theorem is based on a crucial argument, namely the Pitt's inequality \cite{[pitt1], [pitt2], [pitt3]}, it was firstly applied to Landau equations
by Alexandre-Liao-Lin \cite{[Alex]}. For $\ga\in(-2,0)$, we follow Alexandre-Liao-Lin's idea \cite{[Alex]} to construct general $L^{p}$ theorem. However, when $\ga=-2$, it is the critical case, we apply uniform bound of entropy to get a parameter of freedom, it will help us control the coercivity term. Moreover, the entropy production estimate (theorem \ref{theorem3}) plus bootstrap procedure can improve the temporal growth.

\subsection{Application} As an application, we deduce some well-posedness results. Let us recall existence and uniqueness results for the spatially homogeneous Landau equation (\ref{landau}) with $\ga\in (-3,0)$. Consider an initial condition $f_{{\rm in}}(v)\in L^{1}_{2}(\R^{3})\cap L\log L(\R^{3})$, then Fournier-Guerin \cite{[Fournier]} has shown the uniqueness holds in
$$L^{\infty}([0,T],L^{1}_{2})\cap L^{1}([0,T],\mathcal{J}_{\ga})\,,$$
where the space $\mathcal{J}_{\ga}$ of probability measures of $f_{t}$ on $\R^{3}$ such that
$$
J_{\ga}(f_{t})\equiv \sup_{v\in\R^{3}}\int_{\R^{3}}|v-v_{*}|^{\ga}f_{t}(v_{*})dv_{*}<\infty\,.
$$
We have the following estimate about $J_{\ga}(f_{t})$: there exists a constant $C>0$ such that
$$
J_{\ga}(f_{t})\leq \|f_{t}\|_{L^{1}}+C\|f_{t}\|_{L^{p}}\,,
$$
as soon as $p>\frac{3}{3+\ga}$. This implies the uniqueness also holds for
$$L^{\infty}([0,T],L^{1}_{2})\cap L^{1}([0,T],L^{p})\,.$$
For the existence result, Villani \cite{[Villani2]} has shown the existence of weak solution $f_{t}(v)\in L^{\infty}([0,T],L^{1}_{2})$, with constant energy and nonincreasing entropy.

Combining our a priori estimates and existence/uniqueness results listed above, we have the following well-posedness results (compare Fournier-Guerin \cite{[Fournier]}):
\begin{corollary}
For any $\ga\in (-2,0)$, $0<\eps<\frac{3(2+\ga)}{3+\ga}$ and $s>0$. Let the initial condition $f_{{\rm in}}(v)\in L^{1}_{q}(\R^{3})\cap L\log L(\R^{3})$ for some $q=\frac{-3(\ga-s)(2-\eps)}{\eps}>0$. Then the Landau equation \noindent{\rm(\ref{landau})} has an unique weak solution in  $ L^{\infty}([0,T],L^{1}_{2})\cap L^{\alpha}([0,T],L^{3-\eps})$ with $\alpha=\frac{2(3-\eps)}{3(2-\eps)}$. More precisely,
\begin{align*}
\int_{0}^{T}\|f_{t}\|^{\alpha}_{L^{3-\eps}}dt\lesssim (1+T)^{1+\frac{2\eps(3+\eps)}{3(2-\eps)(2+\ga)}}\,.
\end{align*}

\end{corollary}

\begin{corollary}
For any $\ga\in[-2,0)$, $p>\frac{3}{3+\ga}$. Let the initial condition $f_{{\rm in}}(v)\in L^{1}_{2}(\R^{3})\cap L^{p}(\R^{3})$ for $\ga\in (-2,0)$ and $f_{{\rm in}}(v)\in L^{1}_{q}(\R^{3})\cap L^{p}(\R^{3})$ for $\ga=-2$, where $q$ was defined in theorem \ref{theorem3}. Then the Landau equation \noindent{\rm(\ref{landau})} has an unique weak solution in  $ L^{\infty}([0,T],L^{p})$. More precisely,
\begin{align*}
\|f_{t}\|^{p}_{L^{p}}\lesssim C(t)\,,
\end{align*}
where $C(t)\lesssim e^{C t}$ for $\ga\in(-2,0)$ and $C(t)\lesssim \exp\{C t^{z}\}$ for $\ga=-2$, $C>0$ is a positive constant and the polynomial power $z$ was defined in \rm{(\ref{cons})}.
\end{corollary}
\subsection{Preliminaries} In this subsection, we list some well known results of equation (\ref{landau}):
\begin{proposition} \label{prop0}
For any $\ga\in [-2,0)$. Let us consider a weak solution $f_{t}(v)$ of \noindent{\rm(\ref{landau})} with non-negative initial condition
$$
f_{{\rm in}}(v)\in L^{1}_{2}(\R^{3})\cap L\log L(\R^{3})\,,
$$
then
$$
M_{s}(f_{t})<\infty \quad \hbox{for}\quad 0\leq s\leq 2\,, \quad\hbox{and}\quad H(f_{t})<\infty\,.
$$
\end{proposition}
For high order moments estimate, we recall the following result for $\ga\in (-2,0)$ in \cite{[Villani]} (Section 2.4, p.73).
\begin{proposition} \label{prop1}
For any $\ga\in (-2,0)$, $T>0$. Let us consider a weak solution $f_{t}(v)$ of \noindent{\rm(\ref{landau})} with $t\in[0,T]$, assume that the initial condition $f_{{\rm in}}(v)\in L^{1}_{s}(\R^{3})$ , $s>2$, then
$$
M_{s}(f_{t})\lesssim(1+t)\quad \hbox{for}\quad t\in[0,T]\,.
$$
\end{proposition}
If $\ga=-2$, we have the following moments estimate:
\begin{proposition}\label{prop11} If $\ga=-2$, $T>0$. Let us consider a weak solution $f_{t}(v)$ of \noindent{\rm(\ref{landau})} with $t\in [0,T]$, assume that the initial condition $f_{{\rm in}}(v)\in L^{1}_{s}(\R^{3})$ , $s>2$, then
$$
M_{s}(f_{t})\lesssim (1+t)^{(s-2)/3}\quad \hbox{for}\quad t\in[0,T]\,.
$$
\end{proposition}
\noindent{\it Proof.} We still use the abstract notation $\ga$. Let us recall from \cite{[5]} the basic equation for the moments $M_{s}(f_{t})$:
\begin{align*}
\frac{d}{dt}M_{s}(f_{t})&= s\int\int f_{t}(v)f_{t}(v_{*})|v-v_{*}|^{\ga}\big<v\big>^{s-2}\\
&\phantom{xx}{}\times\Big\{ -2|v|^{2}+2|v_{*}|^{2}+(s-2)\Big(\frac{|v|^{2}|v_{*}|^{2}-(v\cdot v_{*})^{2}}{1+|v|^{2}}\Big)       \Big\} dvdv_{*}\,.
\end{align*}
By the H\"older's inequality,
\begin{align*}
&\phantom{xx}{}\int\int f_{t}(v)f_{t}(v_{*})|v-v_{*}|^{\ga}\big<v\big>^{s-2}(1+|v_{*}|^{2}) dvdv_{*}\\
&\leq\Big(\int\int f_{t}(v)f_{t}(v_{*})|v-v_{*}|^{\ga}\big<v\big>^{s} dvdv_{*}\Big)^{(s-2)/s}
\Big(\int\int f_{t}(v)f_{t}(v_{*})|v-v_{*}|^{\ga}(1+|v_{*}|^{2})^{s/2} dvdv_{*}\Big)^{2/s}\\
&=\int\int f_{t}(v)f_{t}(v_{*})|v-v_{*}|^{\ga}\big<v\big>^{s} dvdv_{*}\,,
\end{align*}
and using that fact that
$$
|v|^{2}|v_{*}|^{2}-(v\cdot v_{*})^{2}\leq |v||v_{*}||v-v_{*}|^{2}\,,
$$
we find
\begin{align*}
\frac{d}{dt}M_{s}(f_{t})\leq s(s-2)\int\int f_{t}(v)f_{t}(v_{*})|v-v_{*}|^{\ga+2}\big<v\big>^{s-4}|v||v_{*}|dvdv_{*}\,,
\end{align*}
it is easy to see that if $\ga=-2$,
$$
\frac{d}{dt}M_{s}(f_{t})\leq s(s-2) M_{1}(f_{t})M_{s-3}(f_{t})\lesssim M_{s-3}(f_{t})\,.
$$
This completes the proof of the proposition.
\qed
\bigskip

We need a result on the ellipticity of the matrix $\overline{a}^{f_{t}}$.
\begin{proposition} \label{prop2}
\noindent{\rm(\cite{[Alex], [5]})} For any $\ga\in [-2,0)$, let $C_{0}>0, H_{0}>0$ be two constants, and consider a non-negative function $f_{t}(v)$ such that $M_{s}(f_{t})<C_{0}$, $0\leq s\leq2$, and $H(f_{t})<H_{0}$, then there exists a constant $C_{coer} =C(\ga, C_{0}, H_{0})$ such that
$$
\overline{a}_{ij}^{f_{t}}\xi_{i}\xi_{j}\geq C_{coer} \big<v\big>^{\ga}|\xi|^{2}\,, \quad \hbox{for all} \quad \xi\in \R^{3}\,.
$$
\end{proposition}

Finally, we have the chain rule for the Landau equation in \cite{[5]}:
\begin{proposition} \label{prop3}
Let $f_{t}(v)$ be a weak solution of the Landau equation and $\beta$ be a $C^{1}$ function with $\beta(0)=0$, then at least formally,
\begin{align}\label{beta}
\frac{d}{dt} \int \beta(f_{t}(v))dv=- \int \overline{a}^{f_{t}}\nabla f_{t}(v) \nabla f_{t}(v) \beta''(f_{t}(v))dv- \int \overline{c}^{f_{t}} \phi_{\beta}(f_{t}(v))dv\,,
\end{align}
where
\begin{align*}
\phi'_{\beta}(x)=x\phi''_{\beta}(x)\,,\quad \hbox{and}\quad \phi_{\beta}(0)=0\,.
\end{align*}
\end{proposition}
This property may help us estimate the Landau equation by choosing suitable $\beta$.

\subsection{Plain of the paper}
The paper is organized as follows: we first prove theorem \ref{theorem3} ($L^{\alpha}_{t}L^{3-\eps}_{v}$ estimate) in section 2. Next, we prove theorem \ref{theorem1} ($L^{\infty}_{t}L^{p}_{v}$ estimate) in section 3.

\section{Proof of Theorem \ref{theorem3}}\label{weak-solution}

We now divide the proof of theorem \ref{theorem3} into several steps.
\\
{\em Step 1: Modified entropy.} We apply (\ref{beta}) with $\beta(x)=(x+1)\log(x+1)$, one easily checks that $\beta''(x)=\frac{1}{x+1}$ and $0\leq\phi_{\beta}(x)=x-\log(x+1)\leq x$. Since $H(f_{{\rm in}})<\infty$ by assumption, we easily see that $\int\beta(f_{t}(v))dv<\infty$. The ellipticity of $\overline{a}^{f_{t}}$ in proposition \ref{prop2} immediately implies
\begin{align}\label{W1W2}
&\phantom{xx}{}\frac{d}{dt} \int \beta(f_{t}(v))dv\nonumber\\
&\leq -C_{coer} \int \big<v\big>^{\ga}\frac{|\nabla f_{t}(v)|^{2}}{1+f_{t}(v)}dv+2(\ga+3)\int\int|v-v_{*}|^{\ga}f_{t}(v)f_{t}(v_{*})dvdv_{*}\\
&\equiv W_{1}(f_{t})+W_{2}(f_{t})\nonumber\,.
\end{align}
{\em Step 2: Estimate of $W_{1}(f_{t})$.} First,
\begin{align*}
\int \big<v\big>^{\ga}\frac{|\nabla f_{t}(v)|^{2}}{1+f_{t}(v)}dv&=4\int\big<v\big>^{\ga}|\nabla(\sqrt{1+f_{t}(v)}-1)|^{2}dv\\
&\geq 4(1+D)^{\ga}\|\nabla(\sqrt{1+f_{t}}-1)\|^{2}_{L^{2}(B_{D})}\,,
\end{align*}
for any $D>0$, where $B_{D}=\{ x\in\R^{3}:|x|<D  \}$. By the Sobolev embedding theorem, there exists a constant $C_{sob}>0$ such that
\begin{align*}
\|\nabla(\sqrt{1+f_{t}}-1)\|^{2}_{L^{2}(B_{D})}&\geq C_{sob}\|\sqrt{1+f_{t}}-1\|^{2}_{L^{6}(B_{D})}-\|\sqrt{1+f_{t}}-1\|^{2}_{L^{2}(B_{D})} \\
&\geq C_{sob}\|\sqrt{f_{t}}\textbf{1}_{\{ f_{t}\geq1 \}  }\|^{2}_{L^{6}(B_{D})}-\|f_{t}\|_{L^{1}(B_{D})}\\
&\geq C_{sob}\|f_{t}\textbf{1}_{\{ f_{t}\geq1 \}  }\|_{L^{3}(B_{D})}-M_{0}(f_{t})\\
 &\gtrsim \|f_{t}\textbf{1}_{\{ f_{t}\geq1 \}  }\|_{L^{3}(B_{D})}-1\,.
\end{align*}
Finally, for all $D\geq 1$, since $(1+D)\leq 2D$, we have
\begin{align}\label{W1}
W_{1}(f_{t})\lesssim - D^{\ga}\big(\|f_{t}\textbf{1}_{\{ f_{t}\geq1 \}  }\|_{L^{3}(B_{D})}-1\big)\,.
\end{align}

\textbf{Remark} Basically step 1 and step 2 follow the idea from Desvillettes-Villani \cite{[5]} and Fournier-Guerin \cite{[Fournier]} directly, this is a standard method for entropy production estimate. If we want to improve Fournier-Guerin's result in \cite{[Fournier]}, we need some new ideas in the following steps.
\bigskip

{\em Step 3: Estimate of $W_{2}(f_{t})$.} We shall present the estimates of $W_{2}(f_{t})$ for $\ga\in(-2,0)$ and $\ga=-2$ separately.
\\ \\
{\em Case I: $\ga\in(-2,0)$.} We apply the Hardy-Littlewood-Sobolev inequality \cite{[Lieb]} (Section 4.3, p.106) to get that
$$
W_{2}(f_{t})\leq C_{har}\|f_{t}\|_{L^{q_{1}}}\|f_{t}\|_{L^{q_{2}}}\,,
$$
where $\frac{-\ga}{3}+\frac{1}{q_{1}}+\frac{1}{q_{2}}=2$ or $\frac{1}{q_{1}}+\frac{1}{q_{2}}=\frac{6+\ga}{3}$, if we choose $q_{1}=q_{2}$, then
\begin{align*}
W_{2}(f_{t})\leq C_{har}\|f_{t}\|_{L^\frac{6}{6+\ga}}^{2}\,.
\end{align*}
Let us interpolate $L^\frac{6}{6+\ga}$ between $L^{1}$ and $L^{3-\eps}$, i.e. $\frac{6+\ga}{6}=\frac{x}{3-\eps}+\frac{1-x}{1}$, we have $x=\frac{-\ga(3-\eps)}{6(2-\eps)}$, and hence
\begin{align}\label{W2I}
W_{2}(f_{t})\lesssim\|f_{t}\|_{L^{3-\eps}}^{\frac{-\ga(3-\eps)}{3(2-\eps)}}M_{0}^{\frac{3(4+\ga)-(6+\ga)\eps}{3(2-\eps)}}(f_{t})\lesssim
\|f_{t}\|_{L^{3-\eps}}^{\frac{-\ga(3-\eps)}{3(2-\eps)}}\,.
\end{align}
{\em Case II: $\ga=-2$.}
The estimate of $W_{2}(f_{t})$ for $\ga=-2$ depends on the size of $\|f_{t}\|_{L^{3-\eps}}$. Let $C_{size}$ be any fixed constant. If $\|f_{t}\|_{L^{3-\eps}}<C_{size}$, by (\ref{W2I}),
\begin{align}\label{W2II1}
W_{2}(f_{t})\leq C_{har}\|f_{t}\|_{L^{3-\eps}}^{\frac{2(3-\eps)}{3(2-\eps)}}M_{0}^{\frac{2(3-2\eps)}{3(2-\eps)}}(f_{t})\lesssim 1\,.
\end{align}
On the other hand, if $\|f_{t}\|_{L^{3-\eps}}\geq C_{size}$, we can decompose $W_{2}(f_{t})$ as
\begin{align}\label{W2II2}
W_{2}(f_{t})&= \int\int_{|v-v_{*}|> R_{*}}+ \int\int_{|v-v_{*}|\leq R_{*}}|v-v_{*}|^{-2}f_{t}(v)f_{t}(v_{*})dv dv_{*}\\
&\equiv W_{21}(f_{t})+W_{22}(f_{t})\nonumber\,.
\end{align}
It is easy to see that
\begin{align}
W_{21}(f_{t})\leq R_{*}^{-2}M_{0}^{2}(f_{t})\lesssim R_{*}^{-2}\,.
\end{align}
Next for $W_{22}(f_{t})$, we apply the Cauchy-Schwarz inequality to get that
\begin{align}\label{W22}
W_{22}(f_{t})&\leq\int\int_{|v-v_{*}|\leq R_{*}} \big(|v-v_{*}|^{-1}f_{t}(v)\big)\big(|v-v_{*}|^{-1}f_{t}(v_{*})\big)dv dv_{*}\nonumber\\
&\leq \int\int_{|v-v_{*}|\leq R_{*}} |v-v_{*}|^{-2}f_{t}^{2}(v)dv dv_{*}\\
&\leq R_{*}\int f_{t}^{2}(v)dv\nonumber\,.
\end{align}
Consider the following decomposition:
\begin{align}\label{W221-222}
\int f_{t}^{2}(v)dv&=\int f_{t}^{2}(v){\textbf{1}}_{\{\log f_{t}(v)< A \}}dv+\int f_{t}^{2}(v){\textbf{1}}_{\{\log f_{t}(v)\geq A \}}dv\\
&\equiv W_{221}(f_{t})+W_{222}(f_{t})\nonumber\,,
\end{align}
where $A$ is a parameter to be chosen later. It is easy to see that
\begin{align}\label{W221}
W_{221}(f_{t})=\int f_{t}^{2}(v)\textbf{1}_{\{\log f_{t}(v)< A\}}dv\leq e^{A}M_{0}(f_{t})\lesssim  e^{A}\,.
\end{align}
For $W_{222}(f_{t})$, note that $\log f\geq A$, we can decompose $f_{t}^{2}(v)$ as
$$
f_{t}^{2}(v)\leq\big(f_{t}^{3-\eps}(v)\big)^{1/q_{1}}\big(f_{t}(v)\log f_{t}(v)\big)^{1/q_{2}}\big(f_{t}(v)\big)^{1/q_{2}}A^{-1/q_{2}},
$$
where
$\frac{3-\eps}{q_{1}}+\frac{2}{q_{2}}=2$ and $\frac{1}{q_{1}}+\frac{2}{q_{2}}=1$, this gives $\frac{1}{q_{1}}=\frac{1}{2-\eps}$ and $\frac{1}{q_{2}}=\frac{1-\eps}{2(2-\eps)}$, by the H\"older's inequality,
\begin{align}\label{W222}
W_{222}(f_{t})&=\int f_{t}^{2}(v)\textbf{1}_{\{\log f_{t}(v)\geq A\}}dv\nonumber\\
&\leq A^{\frac{-(1-\eps)}{2(2-\eps)}}\|f_{t}\|^{\frac{3-\eps}{2-\eps}}_{L^{3-\eps}}H^{\frac{1-\eps}{2(2-\eps)}}(f_{t}) M_{0}^{\frac{1-\eps}{2(2-\eps)}}(f_{t})\\
&\lesssim A^{\frac{-(1-\eps)}{2(2-\eps)}}\|f_{t}\|^{\frac{3-\eps}{2-\eps}}_{L^{3-\eps}}\nonumber\,.
\end{align}
Combining (\ref{W2II2})--(\ref{W222}), we have
\begin{align*}
W_{2}(f_{t})&=W_{21}(f_{t})+R_{*}\big(W_{221}(f_{t})+W_{222}(f_{t})\big)\\
&\lesssim R_{*}^{-2}+R_{*}A^{\frac{-(1-\eps)}{2(2-\eps)}}\|f_{t}\|^{\frac{3-\eps}{2-\eps}}_{L^{3-\eps}}+R_{*} e^{A}\,.
\end{align*}
Optimizing the first two terms with respect to $R_{*}$, we obtain
\begin{align}\label{W2II3}
W_{2}(f_{t})&\lesssim A^{\frac{-(1-\eps)}{3(2-\eps)}}\|f_{t}\|^{\frac{2(3-\eps)}{3(2-\eps)}}_{L^{3-\eps}}+ e^{A}A^{\frac{(1-\eps)}{6(2-\eps)}}\|f_{t}\|^{\frac{-(3-\eps)}{3(2-\eps)}}_{L^{3-\eps}}\\
&\lesssim A^{\frac{-(1-\eps)}{3(2-\eps)}}\|f_{t}\|^{\frac{2(3-\eps)}{3(2-\eps)}}_{L^{3-\eps}}+  e^{A}A^{\frac{(1-\eps)}{6(2-\eps)}}\nonumber\,.
\end{align}
This completes the estimate of $W_{2}(f_{t})$ for $\ga=-2$ and $\|f_{t}\|_{L^{3-\eps}}\geq C_{size}$.
All in all, by (\ref{W2II1}) and (\ref{W2II3}), we have obtained, for $\ga=-2$,
\begin{align}\label{W2II4}
W_{2}(f_{t})&\lesssim 1+A^{\frac{-(1-\eps)}{3(2-\eps)}}\|f_{t}\|^{\frac{2(3-\eps)}{3(2-\eps)}}_{L^{3-\eps}}+  e^{A}A^{\frac{(1-\eps)}{6(2-\eps)}}\,.
\end{align}
Combining (\ref{W1W2}), (\ref{W1}), (\ref{W2I}) and (\ref{W2II4}), we have
\begin{align*}
\frac{d}{dt} \int \beta(f_{t}(v))dv\leq -D^{\ga}\big(\|f_{t}\textbf{1}_{\{ f_{t}\geq1 \}  }\|_{L^{3}(B_{D})}-1\big)+
W_{2}(f_{t})\,,
\end{align*}
or
\begin{align}\label{entropy2}
D^{\ga}\int_{0}^{T}\|f_{t}\textbf{1}_{\{ f_{t}\geq1 \}  }\|_{L^{3}(B_{D})}dt\leq C\big(H(f_{{\rm in}})\big)+\int_{0}^{T}W_{2}(f_{t})dt\lesssim
1+\int_{0}^{T}W_{2}(f_{t})dt\,,
\end{align}
where
\begin{equation}\label{in.1.a}
\int_{0}^{T}W_{2}(f_{t})dt\lesssim\left\{\begin{array}{l}
\displaystyle \int_{0}^{T}\|f_{t}\|_{L^{3-\eps}}^{\frac{-\ga(3-\eps)}{3(2-\eps)}}dt\,,\quad \ga\in(-2,0)\,,
\\ \\
\displaystyle \Big(1+e^{A}A^{\frac{(1-\eps)}{6(2-\eps)}}\Big)T+
A^{\frac{-(1-\eps)}{3(2-\eps)}}\int_{0}^{T}\|f_{t}\|^{\frac{2(3-\eps)}{3(2-\eps)}}_{L^{3-\eps}}dt\,, \quad \ga=-2\,.
\end{array}
\right.\end{equation}
{\em Step 4: Weighted estimate.} For $s>0$, we define $g_{t}(v)=\big<v\big>^{\ga-s}f_{t}(v)\textbf{1}_{\{f_{t}(v)\geq1\}}$. Now, we will show the relation between the \emph{weighted function} $\|g_{t}\|_{L^{3}}$ and $\|f_{t}\textbf{1}_{\{ f_{t}\geq1 \}  }\|_{L^{3-\eps}}$. Using (\ref{entropy2}),
\begin{align}\label{weig}
\int_{0}^{T}\|g_{t}\|_{L^{3}}dt&\leq \int_{0}^{T}\sum_{k\geq 0}\|g_{t}\|_{L^{3}(\{2^{k}-1\leq|v|\leq2^{k+1}-1\})}dt\nonumber\\
&\leq  2^{-(\ga-s)}\int_{0}^{T}\sum_{k\geq 0}2^{(k+1)(\ga-s)}\|f_{t}\textbf{1}_{\{ f_{t}\geq1 \}  }\|_{L^{3}(B_{2^{k+1}})}dt\\
&\lesssim  2^{-(\ga-s)}\sum_{k\geq 0}2^{-s(k+1)} \bigg[1+\int_{0}^{T}W_{2}(f_{t})dt\bigg]\nonumber\\
&\lesssim  1+\int_{0}^{T}W_{2}(f_{t})dt\nonumber\,.
\end{align}
On the other hand, let us decompose the function $f_{t}^{3-\eps}(v)$ as
$$
f_{t}^{3-\eps}(v)=\Big(f_{t}(v)\big<v\big>^{\ga-s}\Big)^{3/q_{1}}\Big(f_{t}(v)\big<v\big>^{q}\Big)^{1/q_{2}}\,,
$$
where
$\frac{1}{q_{1}}+\frac{1}{q_{2}}=1$, $\frac{3}{q_{1}}+\frac{1}{q_{2}}=3-\eps$ and $\frac{3(\ga-s)}{q_{1}}+\frac{q}{q_{2}}=0$, then we have
$\frac{1}{q_{1}}=\frac{2-\eps}{2}$, $\frac{1}{q_{2}}=\frac{\eps}{2}$ and $q=\frac{-3(\ga-s)(2-\eps)}{\eps}$, this implies
\begin{align}\label{fres}
\|f_{t}\textbf{1}_{\{ f_{t}\geq1 \}  }\|_{L^{3-\eps}}\leq M^{\frac{\eps}{2}}_{q}(f_{t})\|g_{t}\|^{\frac{1}{\alpha}}_{L^{3}}\,,
\end{align}
where $\alpha=\frac{2(3-\eps)}{3(2-\eps)}$.
\\ \\
{\em Step 5: Conclusion.} Consequently, by (\ref{weig}) and (\ref{fres}),
\begin{align}\label{conclu}
\int_{0}^{T}\|f_{t}\|^{\alpha}_{L^{3-\eps}}dt&\leq \int_{0}^{T}\|f_{t}\textbf{1}_{\{ f_{t}<1 \}  }\|^{\alpha}_{L^{3-\eps}}dt+\int_{0}^{T}\|f_{t}\textbf{1}_{\{ f_{t}\geq1 \}  }\|^{\alpha}_{L^{3-\eps}}dt\nonumber\\
&\leq T M_{0}^{\frac{\alpha}{(3-\eps)}}(f_{T})+\int_{0}^{T}M^{\frac{\eps\alpha}{2}}_{q}(f_{t})\|g_{t}\|_{L^{3}}dt\\
&\lesssim T + M^{\frac{\eps\alpha}{2}}_{q}(f_{T}) \bigg[1+\int_{0}^{T}W_{2}(f_{t})dt\bigg]\nonumber
\,.
\end{align}
We shall present the conclusion for $\ga\in(-2,0)$ and $\ga=-2$ separately.
\\ \\
{\em Case I: $\ga\in(-2,0)$.} Let $\de$ be a parameter to be chosen later, by (\ref{in.1.a}), (\ref{conclu}) and the Young's inequality,
\begin{align*}
\int_{0}^{T}\|f_{t}\|^{\alpha}_{L^{3-\eps}}dt
&\lesssim T + M^{\frac{\eps\alpha}{2}}_{q}(f_{T}) \bigg[ 1+\int_{0}^{T}\|f_{t}\|_{L^{3-\eps}}^{\frac{-\ga(3-\eps)}{3(2-\eps)}}dt\bigg]\\
&\lesssim M^{\frac{\eps\alpha}{2}}_{q}(f_{T}) \bigg[1+\int_{0}^{T}\Big\{\Big(\de\|f_{t}\|^{\frac{-\ga(3-\eps)}{3(2-\eps)}}_{L^{3-\eps}}\Big)^{-2/\ga}+
\de^{\frac{-2}{(2+\ga)}}\Big\}dt\bigg]\,.
\end{align*}
Let $\eta$ be a small number, one can choose $\de$ such that
\begin{align*}
 M^{\frac{\eps\alpha}{2}}_{q}(f_{T}) \de^{-2/\ga}= \eta\quad \hbox{or}\quad  \de=\eta^{-\ga/2} M^{\frac{\ga\eps\alpha}{4}}_{q}(f_{T}) \,.
\end{align*}
This means
\begin{align*}
\int_{0}^{T}\|f_{t}\|^{\alpha}_{L^{3-\eps}}dt
&\lesssim C(T)+\eta \int_{0}^{T}\|f_{t}\|^{\alpha}_{L^{3-\eps}}dt\,,
\end{align*}
where
$$
C(T)\lesssim M^{\frac{\eps\alpha}{2}}_{q}(f_{T})+T\eta^{\frac{\ga}{2+\ga}} M^{\frac{\eps\alpha}{2+\ga}}_{q}(f_{T}) \lesssim T M^{\frac{\eps\alpha}{2+\ga}}_{q}(f_{T})
\lesssim (1+T)^{1+\frac{2\eps(3-\eps)}{3(2-\eps)(2+\ga)}}\,.
$$
\\ \\
{\em Case II: $\ga=-2$.} By (\ref{in.1.a}) and (\ref{conclu}),
\begin{align*}
\int_{0}^{T}\|f_{t}\|^{\alpha}_{L^{3-\eps}}dt
&\lesssim T + M^{\frac{\eps\alpha}{2}}_{q}(f_{T}) \bigg[\Big(1+e^{A}A^{\frac{(1-\eps)}{6(2-\eps)}}\Big)T+
A^{\frac{-(1-\eps)}{3(2-\eps)}}\int_{0}^{T}\|f_{t}\|^{\alpha}_{L^{3-\eps}}dt\bigg]\\
&\lesssim  M^{\frac{\eps\alpha}{2}}_{q}(f_{T}) \bigg[\Big(1+e^{A}A^{\frac{(1-\eps)}{6(2-\eps)}}\Big)T+
A^{\frac{-(1-\eps)}{3(2-\eps)}}\int_{0}^{T}\|f_{t}\|^{\alpha}_{L^{3-\eps}}dt\bigg]\,.
\end{align*}
Let $\eta$ be a small number, one can choose $A$ large enough such that
$$
A^{\frac{-(1-\eps)}{3(2-\eps)}} M^{\frac{\eps\alpha}{2}}_{q}(f_{T}) =\eta\quad \hbox{or} \quad
A=\eta^{\frac{-3(2-\eps)}{1-\eps}} M^{\frac{\eps(3-\eps)}{1-\eps}}_{q}(f_{T}) \,,
$$
then
\begin{align*}
\int_{0}^{T}\|f_{t}\|^{\alpha}_{L^{3-\eps}}dt\lesssim C(T)+\eta\int_{0}^{T}\|f_{t}\|^{\alpha}_{L^{3-\eps}}dt\,,
\end{align*}
where
\begin{align*}
C(T)&\lesssim (1+T)M^{\frac{\eps\alpha}{2}}_{q}(f_{T})A^{\frac{(1-\eps)}{6(2-\eps)}}e^{A}\\
&\lesssim \exp\Big\{ C  \eta^{\frac{-3(2-\eps)}{1-\eps}} M^{\frac{\eps(3-\eps)}{1-\eps}}_{q}(f_{T})    \Big\}\\
&\lesssim\exp\big\{  C T^{z}  \big\}\,,
\end{align*}
for some constant $C>0$ and
$$
z=\frac{(3-\eps)[ 3(2+s)(2-\eps)-2\eps ]}{3(1-\eps)}\,.
$$
This completes the proof of theorem 1. \qed


\section{Proof of Theorem \ref{theorem1}}\label{Fisher-information}
We divide the proof of theorem \ref{theorem1} into several steps.
\\
{\em Step 1: Energy estimate.}
We apply (\ref{beta}) with the function $\beta(x)=\frac{1}{p}x^{p}$,
\begin{align}\label{L2}
\frac{d}{dt}\frac{1}{p}\|f_{t}\|^{p}_{L^{p}}+\mathcal{B}(f_{t})=0\,,
\end{align}
where
\begin{align}\label{B1B2}
\mathcal{B}(f_{t})
&=  (p-1)\int  \int  a(v-v_{*})f_{t}(v_{*}) f^{(p-2)}_{t}(v)\nabla f_{t}(v)\nabla f_{t}(v)dvdv_{*}\nonumber\\
&\phantom{xx}{}-\frac{1}{p} \int  \int  a(v-v_{*})\nabla_{*}f_{t}(v_{*})\nabla f_{t}^{p}(v)dvdv_{*}\\
&\equiv B_{1}(f_{t})+B_{2}(f_{t})\nonumber\,.
\end{align}
More precisely,
\begin{align}\label{B1}
B_{1}(f_{t})= (p-1)\int  \overline{a}^{f_{t}}(v)f^{(p-2)}_{t}(v)|\nabla f_{t}(v)|^{2}dv \,,
\end{align}
and
\begin{align}\label{B2}
B_{2}(f_{t})=-\frac{1}{p} \int  \overline{c}^{f_{t}}(v)f_{t}^{p}(v) dv\,.
\end{align}
{\em Step 2: Estimate of $B_{1}(f_{t})$.}
The ellipticity of $\overline{a}^{f_{t}}$ in proposition \ref{prop2} immediately implies
\begin{align}\label{B1e}
B_{1}(f_{t})&\gtrsim \int  \big<v\big>^{\ga}f^{(p-2)}_{t}(v)|\nabla  f_{t}(v)|^{2}dv\nonumber\\
&= \frac{4}{p^{2}} \|\big<v\big>^{\ga/2}\nabla  f^{p/2}_{t}\|^{2}_{L^{2}}\\
&\gtrsim \big\|\nabla\big(\big<v\big>^{\ga/2}f^{p/2}_{t}\big) \big\|_{L^{2}}^{2}- \|f_{t}\|^{p}_{L^{p}}\nonumber\,.
\end{align}
{\em Step 3: Estimate of $B_{2}(f_{t})$.} We further decompose $B_{2}(f_{t})$ as
\begin{align}\label{B2decom}
B_{2}(f_{t})&=-\frac{1}{p}\int \overline{c}^{f_{t}}(v)f_{t}^{p}(v) dv\nonumber\\
&\lesssim \int\int |v-v_{*}|^{\ga }f_{t}^{p}(v) f_{t}(v_{*})dvdv_{*}\nonumber\\
&= \int\int_{|v-v_{*}|>1} +\int\int_{|v-v_{*}|\leq 1} |v-v_{*}|^{\ga }f_{t}^{p}(v) f_{t}(v_{*})dvdv_{*}\\
&\equiv B_{21}(f_{t})+B_{22}(f_{t})\nonumber\,.
\end{align}
It is easy to see that
\begin{align}\label{B21}
B_{21}(f_{t})\leq M_{0}(f_{t})\|f_{t}\|^{p}_{L^{p}}\lesssim \|f_{t}\|^{p}_{L^{p}}\,.
\end{align}
We shall present the estimates of $B_{22}(f_{t})$ for $\ga\in(-2,0)$ and $\ga=-2$ separately.
\\ \\
{\em Case I: $\ga\in(-2,0)$.} Note that $\big<v_{*}\big>^{\ga}\big<v\big>^{-\ga}\leq C $ if $|v-v_{*}|\leq 1$, we have obtained
\begin{align}\label{newB22}
B_{22}(f_{t})&\leq\int_{v_{*}}\int_{|v-v_{*}|\leq 1} |v-v_{*}|^{\ga}\big<v_{*}\big>^{-\ga}f_{t}(v_{*})\Big(\big<v\big>^{\ga/2}f^{p/2}_{t}(v)\Big)^{2} \big<v_{*}\big>^{\ga}\big<v\big>^{-\ga}dvdv_{*}\nonumber\\
&\lesssim \int_{v_{*}}\big<v_{*}\big>^{-\ga}f_{t}(v_{*})\int_{v} |v-v_{*}|^{\ga}\Big(\big<v\big>^{\ga/2}f^{p/2}_{t}(v) \Big)^{2}dvdv_{*}\,.
\end{align}
We apply the Pitt's inequality \cite{[pitt1],[pitt2],[pitt3]} to get that
$$
\int_{v} |v-v_{*}|^{\ga}\Big(\big<v\big>^{\ga/2}f^{p/2}_{t}(v)\Big)^{2}dv\leq c_{pitt}\int_{\xi} |\xi|^{-\ga}\big|\widehat{\big<v\big>^{\ga/2}f^{p/2}_{t}(v)}\big|^{2}d\xi\\
\,,
$$
then (\ref{newB22}) becomes
\begin{align}\label{newB22-1}
B_{22}(f_{t})&\lesssim M_{-\ga}(f_{t})\int_{\xi} |\xi|^{-\ga}\big|\widehat{\big<v\big>^{\ga/2}f^{p/2}_{t}(v)}\big|^{2}d\xi\nonumber\\
&\lesssim \int_{|\xi|<R_{*}}+\int_{|\xi|\geq R_{*}} |\xi|^{-\ga}\big|\widehat{\big<v\big>^{\ga/2}f^{p/2}_{t}(v)}\big|^{2}d\xi\\
&\lesssim \widetilde{B}_{221}(f_{t})+\widetilde{B}_{222}(f_{t})\nonumber\,,
\end{align}
where $R_{*}$ is a large number to be chosen later. We shall present the estimates of $\widetilde{B}_{221}(f_{t})$ for $1<p<2$ and $p\geq 2$ separately.  If $1<p<2$, by the Parseval's theorem,
\begin{align}\label{newB22-2p}
\widetilde{B}_{221}(f_{t})\lesssim R_{*}^{-\ga}\big\|\widehat{\big<v\big>^{\ga/2}f^{p/2}_{t}(v)}\big\|_{L^{2}}^{2}\lesssim R_{*}^{-\ga}\|f_{t}\|_{L^{p}}^{p}\lesssim R_{*}^{-\ga+3}\|f_{t}\|_{L^{p}}^{p}\,.
\end{align}
If $p\geq 2$, let us interpolate $L^\frac{p}{2}$ between $L^{1}$ and $L^{p}$, we obtain
\begin{align}\label{newB22-2}
\widetilde{B}_{221}(f_{t})\lesssim R_{*}^{-\ga+3}\big\|\widehat{\big<v\big>^{\ga/2}f^{p/2}_{t}(v)}\big\|_{L^{\infty}_{\xi}}^{2}\lesssim R_{*}^{-\ga+3}\|f^{p/2}_{t}\|_{L^{1}}^{2}\lesssim R_{*}^{-\ga+3}\|f_{t}\|_{L^{p}}^{p\frac{(p-2)}{(p-1)}}\,.
\end{align}
This means for all $1<p<\infty$,
\begin{align}\label{newB22-2pp}
\widetilde{B}_{221}(f_{t})\lesssim R_{*}^{-\ga+3}\big(  1+\|f_{t}\|_{L^{p}}^{p}   \big)\,.
\end{align}
For $\widetilde{B}_{222}(f_{t})$, we have
\begin{align}\label{newB22-3}
\widetilde{B}_{222}(f_{t})&\lesssim R_{*}^{-\ga-2}\int_{|\xi|\geq R_{*}} |\xi|^{2}\big|\widehat{\big<v\big>^{\ga/2}f^{p/2}_{t}(v)}\big|^{2}d\xi
\lesssim R_{*}^{-\ga-2}\big\|\nabla\big(\big<v\big>^{\ga/2}f^{p/2}_{t}\big) \big\|_{L^{2}}^{2}\,.
\end{align}
Combining (\ref{newB22-2pp})--(\ref{newB22-3}), we get
\begin{align}\label{newB22-4}
B_{22}(f_{t})&\lesssim R_{*}^{-\ga+3}\big(  1+\|f_{t}\|_{L^{p}}^{p}   \big)+R_{*}^{-\ga-2}\big\|\nabla\big(\big<v\big>^{\ga/2}f^{p/2}_{t}\big) \big\|_{L^{2}}^{2}\,.
\end{align}
Using (\ref{B21}) and (\ref{newB22-4}), we have the estimate for $B_{2}(f)$,
\begin{align}\label{newB22-5}
B_{2}(f_{t})&\lesssim\|f_{t}\|^{p}_{L^{p}}+ R_{*}^{-\ga+3}\big(  1+\|f_{t}\|_{L^{p}}^{p}   \big)+R_{*}^{-\ga-2}\big\|\nabla\big(\big<v\big>^{\ga/2}f^{p/2}_{t}\big) \big\|_{L^{2}}^{2}\,.
\end{align}
\\
{\em Case II: $\ga=-2$.} We further decompose $B_{22}(f_{t})$ as
\begin{align}\label{newB22decom}
B_{22}(f_{t})&=\int\int_{|v-v_{*}|\leq 1} |v-v_{*}|^{-2 }f_{t}^{p}(v) f_{t}(v_{*}) \big\{{\textbf{1}}_{\{\log f_{t}(v_{*})\geq A \}}+{\textbf{1}}_{\{\log f_{t}(v_{*})< A \}} \big\}dvdv_{*}\\
&\equiv B_{221}(f_{t})+B_{222}(f_{t})\nonumber\,,
\end{align}
where $A=A(t)$ is a large parameter to be chosen later. It is easy to see that
\begin{align}\label{B222}
B_{222}(f_{t})\leq e^{A}\|f_{t}\|^{p}_{L^{p}}\,.
\end{align}
For the difficult part $B_{221}(f_{t})$, since $\big<v_{*}\big>^{-2}\big<v\big>^{2}\leq 3 $, we get
\begin{align*}
B_{221}(f_{t})&\leq\int_{v_{*}}\int_{|v-v_{*}|\leq 1} |v-v_{*}|^{-2}\big<v_{*}\big>^{2}f_{t}(v_{*}){\textbf{1}}_{\{\log f_{t}(v_{*})\geq A \}}\Big(\big<v\big>^{-1}f^{p/2}_{t}(v)\Big)^{2} \big<v_{*}\big>^{-2}\big<v\big>^{2}dvdv_{*}\\
&\lesssim \int_{v_{*}}\big<v_{*}\big>^{2}f_{t}(v_{*}){\textbf{1}}_{\{\log f_{t}(v_{*})\geq A \}}\int_{v} |v-v_{*}|^{-2}\Big(\big<v\big>^{-1}f^{p/2}_{t}(v) \Big)^{2}dvdv_{*}\,.
\end{align*}
We apply the Pitt's inequality to get that
$$
\int_{v} |v-v_{*}|^{-2}\Big(\big<v\big>^{-1}f^{p/2}_{t}(v)\Big)^{2}dv\leq c_{pitt}\int_{\xi} |\xi|^{2}\big|\widehat{\big<v\big>^{-1}f^{p/2}_{t}(v)}\big|^{2}d\xi\,,
$$
then
\begin{align}\label{B22-1}
B_{221}(f_{t})&\lesssim\int_{v_{*}}\big<v_{*}\big>^{2}f_{t}(v_{*}){\textbf{1}}_{\{\log f_{t}(v_{*})\geq A \}}dv_{*}
\int_{\xi} |\xi|^{2}\big|\widehat{\big<v\big>^{-1}f^{p/2}_{t}(v)}\big|^{2}d\xi\\
&\equiv B_{2211}(f_{t})B_{2212}(f_{t})\nonumber\,.
\end{align}
We now analyze the term $B_{2211}(f_{t})$, note that $\log f_{t}(v_{*})\geq A$, we can decompose $\big<v_{*}\big>^{2}f_{t}(v_{*})$ as
$$
\big<v_{*}\big>^{2}f_{t}(v_{*})\leq \big(\big<v_{*}\big>^{4}f_{t}(v_{*})\big)^{1/2}\big(f_{t}(v_{*})\log f_{t}(v_{*})\big)^{1/2}A^{-1/2},
$$
hence the H\"older's inequality implies
\begin{align}\label{B22-3}
B_{2211}(f_{t})\leq A^{-1/2}H^{1/2}(f_{t})M_{4}^{1/2}(f_{t})\lesssim A^{-1/2}M_{4}^{1/2}(f_{t})\,.
\end{align}
Next for $B_{2212}(f_{t})$,
\begin{align}\label{B22-2}
B_{2212}(f_{t})=\big\|\nabla\big(\big<v\big>^{-1}f^{p/2}_{t}\big) \big\|_{L^{2}}^{2}\,.
\end{align}
Combining (\ref{B21}), (\ref{B222}) and (\ref{B22-1})--(\ref{B22-2}), we have
\begin{align}\label{B22-4}
B_{2}(f_{t})&\lesssim A^{-1/2}M_{4}^{1/2}(f_{t})\big\|\nabla\big(\big<v\big>^{-1}f^{p/2}_{t}\big) \big\|_{L^{2}}^{2}+\|f_{t}\|^{p}_{L^{p}}+e^{A}\|f_{t}\|^{p}_{L^{p}}\,.
\end{align}
{\em Step 4: Conclusion.} We shall present the conclusion for $\ga\in(-2,0)$ and $\ga=-2$ separately.
\\ \\
{\em Case I: $\ga\in(-2,0)$.} Consequently, by (\ref{L2}), (\ref{B1B2}), (\ref{B1e}) and (\ref{newB22-5})
\begin{align}\label{newB2con}
\frac{d}{dt}\|f_{t}\|^{p}_{L^{p}}+(1-R_{*}^{-\ga-2}) \big\|\nabla\big(\big<v\big>^{\ga/2}f^{p/2}_{t}\big) \big\|_{L^{2}}^{2}\lesssim R_{*}^{-\ga+3}\big(1+\|f_{t}\|_{L^{p}}^{p}\big)\,.
\end{align}
One can choose $R_{*}$ lagre enough, then it is easy to get
$$
\|f_{t}\|^{p}_{L^{p}}\lesssim e^{C t}\,,
$$
for some constant $C>0$.
\\ \\
{\em Case II: $\ga=-2$.}
By (\ref{L2}), (\ref{B1B2}), (\ref{B1e}) and (\ref{B22-4})
\begin{align}\label{B2con}
&\phantom{xx}{}\frac{d}{dt}\|f_{t}\|^{p}_{L^{p}}+ \big\|\nabla\big(\big<v\big>^{-1}f^{p/2}_{t}\big) \big\|_{L^{2}}^{2}\nonumber\\
&\lesssim A^{-1/2}M_{4}^{1/2}(f_{t})\big\|\nabla\big(\big<v\big>^{-1}f^{p/2}_{t}\big) \big\|_{L^{2}}^{2}+\|f_{t}\|^{p}_{L^{p}}+e^{A}\|f_{t}\|^{p}_{L^{p}}\,.
\end{align}
Let us prove the case $1<p\leq 2$ first. We need more analysis for $e^{A}\|f_{t}\|^{2}_{L^{2}}$. Consider the following decomposition:
$$
f_{t}^{p}(v)=\big(\big<v\big>^{-1}f^{p/2}_{t}(v)\big)^{6/q_{1}}\big(f_{t}^{3-\eps}(v)\big)^{1/q_{2}}\big(\big<v\big>^{d}f_{t}(v)\big)^{1/q_{3}},
$$
where
$\frac{3p}{q_{1}}+\frac{3-\eps}{q_{2}}+\frac{1}{q_{3}}=p$, $\frac{1}{q_{1}}+\frac{1}{q_{2}}+\frac{1}{q_{3}}=1$ and $\frac{-6}{q_{1}}+\frac{d}{q_{3}}=0$, one can choose $\frac{1}{q_{1}}=\frac{p-1}{3p}$, $\frac{1}{q_{2}}=\frac{p-1}{3p(2-\eps)}$, $\frac{1}{q_{3}}=\frac{3(p+1)-(2p+1)\eps}{3p(2-\eps)}$ and $d=\frac{6(2-\eps)(p-1)}{3(p+1)-(2p+1)\eps}<2$, by the H\"older's inequality,
\begin{align*}
\|f_{t}\|^{p}_{L^{p}}&\leq  \|\big<v\big>^{-1} f^{p/2}_{t}\|^{\frac{2(p-1)}{p}}_{L^{6}} \| f_{t}\|^{\frac{(p-1)(3-\eps)}{3p(2-\eps)}}_{L^{3-\eps}}M^{\frac{3(p+1)-(2p+1)\eps}{3p(2-\eps)}}_{d}(f_{t})\\
&\lesssim \big\|\nabla\big(\big<v\big>^{-1}f^{p/2}_{t}\big) \big\|^{\frac{2(p-1)}{p}}_{L^{2}}   \| f_{t}\|^{\frac{(p-1)(3-\eps)}{3p(2-\eps)}}_{L^{3-\eps}}\,,
\end{align*}
and hence by the Young's inequality, there exists a small number $\eta_{1}>0$ such that
\begin{align}\label{B23r2}
e^{A}\|f_{t}\|^{p}_{L^{p}}&\lesssim e^{A} \| f_{t}\|^{\frac{(p-1)(3-\eps)}{3p(2-\eps)}}_{L^{3-\eps}}\big\|\nabla\big(\big<v\big>^{-1}f^{p/2}_{t}\big) \big\|^{\frac{2(p-1)}{p}}_{L^{2}}  \nonumber\\
&\lesssim \eta^{\frac{-(p-1)}{(3-p)}}_{1}e^{\frac{2p}{3-p}A}+\eta_{1}^{-1}\| f_{t}\|^{\frac{2(3-\eps)}{3(2-\eps)}}_{L^{3-\eps}}+\eta_{1}\big\|\nabla\big(\big<v\big>^{-1}f^{p/2}_{t}\big) \big\|^{2}_{L^{2}}  \,.
\end{align}
Combining (\ref{B2con}) and (\ref{B23r2}), we have
\begin{align}\label{B2con12}
&\phantom{xx}{}\frac{d}{dt}\|f_{t}\|^{p}_{L^{p}}+(1-\eta_{1}) \big\|\nabla\big(\big<v\big>^{-1}f^{p/2}_{t}\big) \big\|_{L^{2}}^{2}\nonumber\\
&\lesssim A^{-1/2}M_{4}^{1/2}(f_{t})\big\|\nabla\big(\big<v\big>^{-1}f^{p/2}_{t}\big) \big\|_{L^{2}}^{2}+\|f_{t}\|^{p}_{L^{p}}+ \eta^{\frac{-(p-1)}{(3-p)}}_{1}e^{\frac{2p}{3-p}A}+\eta_{1}^{-1}\| f_{t}\|^{\alpha}_{L^{3-\eps}}\,.
\end{align}
Let $\eta_{2}>0$ be a small number, one can choose $A=A(t)$ large enough such that
$$
 A^{-1/2}M_{4}^{1/2}(f_{t})=\eta_{2}\,,
$$
or say
$$
A(t)=\eta_{2}^{-2}M_{4}(f_{t})\lesssim \eta_{2}^{-2}t^{2/3}\,,
$$
thus (\ref{B2con12}) can be rewritten as
\begin{align*}
&\phantom{xx}{}\frac{d}{dt}\|f_{t}\|^{p}_{L^{p}}+(1 -\eta_{1}-\eta_{2}) \big\|\nabla\big(\big<v\big>^{-1}f^{p/2}_{t}\big) \big\|_{L^{2}}^{2}
\\
 &\lesssim \|f_{t}\|^{p}_{L^{p}}+  \eta^{\frac{-(p-1)}{(3-p)}}_{1}\exp\Big\{\frac{2p}{(3-p)}\eta_{2}^{-2}t^{2/3}  \Big\}+\eta_{1}^{-1}\| f_{t}\|^{\alpha}_{L^{3-\eps}}\,,
\end{align*}
using the estimate of theorem \ref{theorem3}, we have
\begin{align*}
\|f_{t}\|^{2}_{L^{2}}\lesssim C(t)\,,
\end{align*}
where
$$
C(t)\lesssim \exp\{C t^{z}\}\,,
$$
$C>0$ is a positive constant and the polynomial power $z$ was defined in (\ref{cons}). This completes the case $1<p\leq 2$. For $p>2$, we will prove the following argument: Let $p\geq 2, 0<\de\leq 1$ and the initial condition $f_{in}\in L^{p+\de}$, if $\|f_{t}\|^{p}_{L^{p}}\lesssim \exp\{C t^{z}\}$, then $\|f_{t}\|^{p+\de}_{L^{p+\de}}\lesssim \exp\{C' t^{z}\}$, where $C\leq C'$. If one can prove this argument, then the bootstrap procedure finishes the proof of theorem.
\\ \\
\noindent{\it Proof of the argument.} For the case $p+\de$, note that (\ref{B2con}) can be rewritten as
\begin{align}\label{B2conp1}
&\phantom{xx}{}\frac{d}{dt}\|f_{t}\|^{p+\de}_{L^{p+\de}}+ \big\|\nabla\big(\big<v\big>^{-1}f^{(p+\de)/2}_{t}\big) \big\|_{L^{2}}^{2}\nonumber\\
&\lesssim A^{-1/2}M_{4}^{1/2}(f_{t})\big\|\nabla\big(\big<v\big>^{-1}f^{(p+\de)/2}_{t}\big) \big\|_{L^{2}}^{2}+\|f_{t}\|^{p+\de}_{L^{p+\de}}+e^{A}\|f_{t}\|^{p+\de}_{L^{p+\de}}\,.
\end{align}
Again, we need more analysis for $e^{A}\|f_{t}\|^{p+\de}_{L^{p+\de}}$. Consider the following decomposition:
$$
f_{t}^{p+\de}(v)=\big(\big<v\big>^{-1}f^{(p+\de)/2}_{t}(v)\big)^{6/q_{1}}\big(f_{t}^{p}(v)\big)^{1/q_{2}}
\big(\big<v\big>^{d}f_{t}(v)\big)^{1/q_{3}},
$$
where
$\frac{3(p+\de)}{q_{1}}+\frac{p}{q_{2}}+\frac{1}{q_{3}}=p+\de$, $\frac{1}{q_{1}}+\frac{1}{q_{2}}+\frac{1}{q_{3}}=1$ and $\frac{-6}{q_{1}}+\frac{d}{q_{3}}=0$, one can choose $\frac{1}{q_{1}}=\frac{1}{6}$, $\frac{1}{q_{2}}=\frac{3(p+\de)-5}{6(p-1)}$, $\frac{1}{q_{3}}=\frac{2p-3\de}{6(p-1)}$ and $d=q_{3}$, by the H\"older's inequality, there exists small number $\eta_{1}>0$ such that
\begin{align}\label{p1}
e^{A}\|f_{t}\|^{p+\de}_{L^{p+\de}}&\leq  e^{A}\|\big<v\big>^{-1} f^{(p+\de)/2}_{t}\|_{L^{6}} \|f_{t}\|^{\frac{p}{q_{2}}}_{L^{p}} M^{\frac{1}{q_{3}}}_{d}(f_{t})\nonumber\\
&\lesssim
\eta_{1}\big\|\nabla\big(\big<v\big>^{-1}f^{(p+\de)/2}_{t}\big) \big\|^{2}_{L^{2}}+\eta_{1}^{-1}  e^{2A}
\|f_{t}\|^{\frac{2p}{q_{2}}}_{L^{p}} M^{\frac{2}{q_{3}}}_{d}(f_{t})\\
&\lesssim
\eta_{1}\big\|\nabla\big(\big<v\big>^{-1}f^{(p+\de)/2}_{t}\big) \big\|^{2}_{L^{2}}+\eta_{1}^{-1}  \exp\{2A+C t^{z}\}
\nonumber\,,
\end{align}
similar to $1<p\leq 2$, let $\eta_{2}>0$ be a small number, one can choose $A=A(t)$ large enough such that
\begin{align}\label{Ap1}
 A^{-1/2}M_{4}^{1/2}(f_{t})=\eta_{2}\,.
\end{align}
Combining (\ref{B2conp1})--(\ref{Ap1}), we have
\begin{align}\label{B2con1}
\frac{d}{dt}\|f_{t}\|^{p+\de}_{L^{p+\de}}+(1-\eta_{1}-\eta_{2}) \big\|\nabla\big(\big<v\big>^{-1}f^{(p+\de)/2}_{t}\big) \big\|_{L^{2}}^{2}\lesssim \exp\{\eta_{2}^{-2}t^{2/3}+C t^{z}\}\,.
\end{align}
hence
\begin{align*}
\|f_{t}\|^{(p+\de)}_{L^{(p+\de)}}\lesssim C(t)\,,
\end{align*}
where
$$
C(t)\lesssim \exp\{C't^{z}\}\,,
$$
the polynomial power $z$ was defined in (\ref{cons}). This completes the proof of theorem 2.

 \qed


\end{document}